\documentstyle[preprint,psfig]{aastex}

\def\lesssim{\mathrel{\hbox{\rlap{\hbox{\lower4pt\hbox{$\sim$}}}\hbox{$<$}}}}
\def\gtrsim{\mathrel{\hbox{\rlap{\hbox{\lower4pt\hbox{$\sim$}}}\hbox{$>$}}}}

\begin{document}

\title{High Resolution Simulations of the Plunging Region in a
Pseudo-Newtonian Potential:  Dependence on Numerical Resolution
and Field Topology}

\author{John F. Hawley}
\affil{Department of Astronomy, University of Virginia,
Charlottesville VA 22903}

\author{Julian H. Krolik}
\affil{Physics and Astronomy Department, Johns Hopkins University,
    Baltimore, MD 21218}

\shorttitle{Global MHD Simulation II}

\begin{abstract}

New three dimensional magnetohydrodynamic simulations of accretion disk
dynamics in a pseudo-Newtonian Paczy\'nski-Wiita potential are
presented.  These have finer resolution in the inner disk than any
previously reported.  Finer resolution leads to increased magnetic
field strength, greater accretion rate, and greater fluctuations in the
accretion rate.  One simulation begins with a purely poloidal magnetic
field, the other with a purely toroidal field.  Compared to the
poloidal initial field simulation, a purely toroidal initial field
takes longer to reach saturation of the magnetorotational instability
and produces less turbulence and weaker magnetic field energies.  For
both initial field configurations, magnetic stresses continue across
the marginally stable orbit; measured in units corresponding to the
Shakura-Sunyaev $\alpha$ parameter, the stress grows from $\sim 0.1$ in
the disk body to as much as $\sim 10$ deep in the plunging region.
Matter passing the inner boundary of the simulation has $\sim 10\%$
greater binding energy and $\sim 10\%$ smaller angular momentum than it
did at the marginally stable orbit.  Both the mass accretion rate and
the integrated stress fluctuate widely on a broad range of timescales.

\end{abstract}


\section{Introduction}

MHD turbulence driven by magneto-rotational instability (MRI; Balbus \&
Hawley 1991) now appears to be the fundamental physical mechanism of
angular momentum transport in accretion disks (see the review by Balbus
\& Hawley 1998).  On the basis of this idea, it is now possible to
begin to answer many questions about accretion dynamics.  In a series
of papers, we are investigating the global radial structure of
accretion disks near the marginally stable orbit of a black hole by
means of large-scale three dimensional MHD numerical simulations.  In
particular, we focus on the inner regions of accretion disks, and
examine the time-dependence of accretion, the radial dependence of
stress and dissipation, and the net energy and angular momentum per
unit mass carried into the black hole.  In so doing we will also need
to examine more closely the conceptual basis for disk dynamics.

For understandable reasons, simple analytic models (e.g., Novikov
\& Thorne 1973, Shakura \& Sunyaev 1973) have in general been
built on the assumption of time-steadiness.  However, this assumption
is by no means a given in real disks.  Even if there is a long-term
mean accretion rate, it is entirely possible for there to be short-term
fluctuations.  In fact, sizable fluctuations seen in the light-curves
of every accreting black hole indicate that accretion variability is
the norm, not the exception (e.g., Sunyaev \& Revnivtsev 2000).  

The difficulty with addressing issues of time-dependent dynamics within
the context of traditional analytic models is that the dynamics of
accretion depend fundamentally on the nature of the stress that
transports angular momentum.  While many steady-state or time-averaged
properties of disks may be adequately described by a simple stress
parameterization (e.g., the Shakura-Sunyaev $\alpha$ prescription in
which the stress is supposed proportional to pressure), the actual
dynamics cannot be so treated.  $\alpha$ is only a measure of the
stress; it is not the physics behind the stress.  Within the $\alpha$
parameterization, stress results from turbulent fluctuations ({\it not}
viscosity).  These fluctuations have amplitudes less than or of order
the sound speed $c_s$ on scales less than or of order the disk
scale-height $H$.  Again, while there may be regions of the disk where
it is sufficient to time-average over these fluctuations, this cannot
be the case near the black hole where accretion time- and length-scales
become comparable to those that characterize the turbulence.  Direct
dynamical simulations are required to understand time-dependent
quantities.

Although it has long been clear that ordinary viscosity cannot account
for angular momentum transport in accretion disks, modeling the
stress ``as if" it were viscous has been a popular way of thinking
about disks for an equally long time.  Again, for
dynamical issues this is clearly incorrect:  a low-viscosity plasma
that is turbulent does not behave like high-viscosity laminar flow.  A
separate question is whether the turbulent stress would behave
sufficiently similarly to viscosity that there would be an associated
dissipation rate proportional to the local stress (Novikov \& Thorne
1973; Shakura \& Sunyaev 1973).  Not all stresses have this property,
although {\it if} the MRI-driven MHD turbulence dissipates locally in a
turbulent cascade it is amenable to an $\alpha$-type description
(Balbus \& Papaloizou 1999).  Because we now know that the dominant
stress is electromagnetic, local dissipation is not certain; Poynting
flux can easily carry energy from one place to another, and
highly-magnetized coronae and winds may account for much of the
liberated energy.  Numerical simulations offer the possibility of
measuring to what degree this happens.

Both the radial dependence of stress and dissipation and the net energy
and angular momentum delivered to the black hole depend on yet another
plausible, but not demonstrated, assumption: that the inter-ring stress
disappears in the vicinity of the marginally stable orbit.  Two
heuristic arguments were raised in behalf of this assumption: that the
small amount of mass in the plunging region could hardly be expected to
exert a force on the far heavier disk proper (Novikov \& Thorne 1973,
Page \& Thorne 1974); and if the stress scales as a constant fraction
$\alpha$ of the local pressure, then the low pressure in the plunging
region would lead to a very small stress (Abramowicz \& Kato 1989).
However, as recognized by Page \& Thorne (1974), neither of these
arguments applies to magnetic stresses.  Krolik (1999) and Gammie
(1999) argued that the dominant role of magnetic stresses in angular
momentum transport in the disk body should actually lead to stresses
near the marginally stable orbit large enough to substantially alter
the amount of energy and angular momentum removed from matter before it
passes through the black hole's event horizon.  If so, the radial
distribution of stress in the disk would also be significantly altered,
with wide-ranging observational consequences (Agol \& Krolik 2000).
Simulations can quantitatively evaluate the importance of this mechanism.

Global three-dimensional disk simulations have only recently become
possible, and the number of such models is still sufficiently small
that it remains possible to give a nearly comprehensive list of
references:  Armitage (1998); Matsumoto (1999); Hawley (2000, hereafter
H00); Machida et al.\ (2000); Hawley \& Krolik (2001, hereafter HK01);
Hawley (2001); Armitage et al.\  (2001).  All of this work has shared
two key assumptions:  Newtonian dynamics in a Newtonian or
pseudo-Newtonian (Paczy\'nski-Wiita) potential, and a fixed (adiabatic
or isothermal) equation of state.  All of this work has also struggled
with the same central problem:  obtaining resolution adequate to
describing the physics.

In this paper, we report simulations with the best resolution in the
inner accretion flow yet achieved.  We also use these simulations to
explore whether the topology of the initial seed magnetic field has any
lasting effects on the structure of the accretion flow.  In later
efforts we will improve the level of realism in these simulations by
solving the energy equation and employing genuine relativistic
dynamics.

\section{Numerical Method}
As in past work (H00; HK01) we evolve the
equations of Newtonian MHD in cylindrical coordinates $(R,\phi,z)$,
namely
\begin{equation}\label{mass}
{\partial\rho\over \partial t} + \nabla\cdot (\rho {\bf v}) =  0
\end{equation}
\begin{equation}\label{mom}
\rho {\partial{\bf v} \over \partial t}
+ (\rho {\bf v}\cdot\nabla){\bf v} = -\nabla\left(
P + {\mathcal Q} +{B^2\over 8 \pi} \right)-\rho \nabla \Phi +
\left( {{\bf B}\over 4\pi}\cdot \nabla\right){\bf B}
\end{equation}
\begin{equation}\label{ene}
{\partial\rho\epsilon\over \partial t} + \nabla\cdot (\rho\epsilon
{\bf v}) = -(P+{\mathcal Q}) \nabla \cdot {\bf v}
\end{equation}
\begin{equation} \label{ind}
{\partial{\bf B}\over \partial t} =
\nabla\times\left( {\bf v} \times {\bf B} \right)
\end{equation}
where $\rho$ is the mass density, $\epsilon$ is the specific internal
energy, ${\bf v}$ is the fluid velocity, $P$ is the pressure, $\Phi$
is the gravitational potential, ${\bf B}$ is the magnetic field vector,
and ${\mathcal Q}$ is an explicit artificial viscosity of
the form described by Stone \& Norman (1992a).
To model a black hole gravitational field we use
the pseudo-Newtonian potential of Paczy\'nski \& Wiita (1980) which is
\begin{equation}\label{pwp}
\Phi = - {G M \over r-r_g},
\end{equation}
where $r$ is spherical radius,
and $r_g \equiv 2GM/c^2$ is the ``gravitational radius,''
akin to the black hole horizon.   For this potential,
the Keplerian specific angular momentum (i.e., that
corresponding to a circular orbit) is
\begin{equation}\label{pwl}
l_{kep} = (GMr)^{1/2} {r \over r-r_g} ,
\end{equation}
and the angular frequency $\Omega = l/R^2$.  The innermost
marginally stable circular orbit is located at $r_{ms}=3r_g$.  We use
an adiabatic equation of state, $P=\rho\epsilon(\Gamma -1) =
K\rho^\Gamma$, where $P$ is the pressure, $\rho$ is the mass density,
$\epsilon$ is the specific internal energy, $K$ is a constant, and
$\Gamma = 5/3$.  Radiation transport and losses are omitted.  Since
there is no explicit resistivity or physical viscosity, the gas can
heat only through adiabatic compression or by artificial viscosity
which acts in shocks.

The code employs time-explicit Eulerian finite differencing.  The
numerical algorithm is that of the ZEUS code for hydrodynamics (Stone
\& Norman 1992a) and MHD (Stone \& Norman 1992b; Hawley \& Stone
1995).  We set $GM=1$ and $r_g = 1$ (so that $c = \sqrt{2}$), thus
establishing the units of time and velocity.  The circular orbital
period at a radius $r$ is $P_{orb} = 2\pi \Omega^{-1} = 2\pi r^{3/2}
(r-r_g)/r$.

In this paper we increase the overall resolution within the disk itself
and in the inflow region above what was used in H00 and HK01.  The
computational grid is laid out in cylindrical coordinates, with
$256\times 64 \times 192$ zones in $R \times \phi \times z$.  This
represents only a 50\% increase in the total number of zones used
compared to HK01.  In the present simulations, however, the zones are
concentrated to increase the effective resolution in the most important
regions of the flow.  We locate more of the zones near the marginally
stable orbit and around the equator, and double the angular resolution
while decreasing the angular extent.  The radial inner boundary is
moved in to $R_{\rm min} =1.25$ and there are 110 equally spaced zones
out to $R=4$.  Compared to our earlier simulation, this scheme
decreases the zone size $\Delta R$ in the inner region by a factor of
3.3.  Beyond $R=4$, $\Delta R$ gradually increases; the remaining 146
zones extend out to $R=36$.  The $z$ coordinate is centered on the
equatorial plane, and runs from -11 to $+11$.  From $z=-1$ to $1$ there
are 76 equally spaced zones; again comparing to the earlier simulation,
the $\Delta z$ around the equator is smaller by a factor of 2.4.
Beyond $z=\pm 1$, $\Delta z$ gradually increases out to the top and
bottom boundaries.

The angle $\phi$ spans the range from 0 to $\pi/2$ in 64 equally spaced
zones;  $\Delta \phi$ is half the size used in H00 and HK01.  Although
the resolution is improved over H00 and HK01, the domain is only one
quarter as large.  However, experiments (Hawley 2001) with
``cylindrical'' disks (no vertical gravity) found that reducing the
angular domain from $2\pi$ to $\pi/2$ does not alter the qualitative
features of the evolution, although it lowered the energy and stress
levels by about 10\%.  Since the practical advantage of limiting the
angular domain is great, we use it here and assume that the
quantitative effects will be small.

The boundary conditions on the grid are simple zero-gradient outflow
conditions; no flow into the computational domain is permitted.  The
magnetic field boundary condition is set by requiring the transverse
components of the field to be zero outside the computational domain,
while the perpendicular component satisfies the divergence-free
constraint.  The $\phi$ direction is, of course, periodic.

The initial condition for the simulations is the same torus used in
HK01 and for model GT4 of H00.  This is a moderately thick torus
($H/R \simeq 0.12$ at the pressure maximum) with an angular velocity
distribution $\Omega \propto R^{-1.68}$, slightly steeper than
Keplerian.  The angular momentum within the torus is equal to the
Keplerian value at the torus pressure maximum at $R=10$.  As before,
the pressure and density at $R=10$ are $P_{max}=0.036$ and
$\rho_{max}=34$, while $P_{orb}(R=10)=179$.  For reference, the orbital
period at the marginally stable orbit is $P_{orb} =21.8$.

We consider two different initial field configurations:  poloidal
loops, as in HK01 and H00, and a purely toroidal field.  These models
are discussed in turn in \S3 and \S 4.

\section{Initially Poloidal Magnetic Field}

   The initial magnetic field topology for the the first of the two
simulations reported here is identical to that of the GT4 simulation
in H00 and the simulation
presented in HK01: poloidal field loops lying along equal density
surfaces in the torus.  The initial condition for the magnetic field is
set by the toroidal component of the vector potential $A_\phi (R,z) =
\rho (R,z) - \rho_{min}$, for all $\rho$ greater than a minimum value,
$\rho_{min} = 0.1$.  The poloidal field is then obtained from ${\bf B}
= \nabla \times {\bf A}$.  The resulting field is fully contained
within the torus.  The strength of the magnetic field is set so that
the ratio of the total integrated gas pressure to magnetic pressure is
100, i.e., $\beta=P_{\rm gas}/P_{\rm mag} = 100$.

   Only three differences distinguish this simulation from the previous
ones:  the earlier simulations treated a full $2\pi$ in azimuth,
whereas this one considers only a single quadrant; the gridding scheme
used in the new one offers an improvement of roughly a factor of 3 in
$R$ and $z$ resolution, and a factor of 2 in $\phi$ resolution; and the
new simulation was run somewhat longer, to $t=1838$ rather than
$t=1500$.  This longer time corresponds to 84 orbits at $r_{ms}$, and
is reached in 560,000 timesteps.  Two purposes are served by studying
this new simulation: we examine the degree to which the higher
resolution of the new simulation allows us to approach numerical
convergence, and it provides a standard of comparison for the
simulation reported in the next section whose initial magnetic field
was purely toroidal.

     Just as in earlier simulations, the poloidal field loops of the
initial state have unstable MRI wavelengths that are well-resolved on
the grid.  As a result, the MRI grows rapidly and turbulence develops
within the disk.  The disk expands as the MHD turbulence redistributes
angular momentum.  When the inner edge reaches the marginally stable
orbit, matter begins to accrete through the inner boundary of the
simulation, into the ``black hole".  The initial growth phase of total
field energy ends at $t \simeq 600$.  For the remainder of the
simulation, conditions in the disk exhibit a rough stationarity, but
with very large fluctuations in almost every quantity.
The disk is modestly thick with $H/R \simeq 0.21$ at $R=10$ and
$H/R \simeq 0.15$ at $R=3$.  

\begin{figure}[htb]
\centerline{
\psfig{file=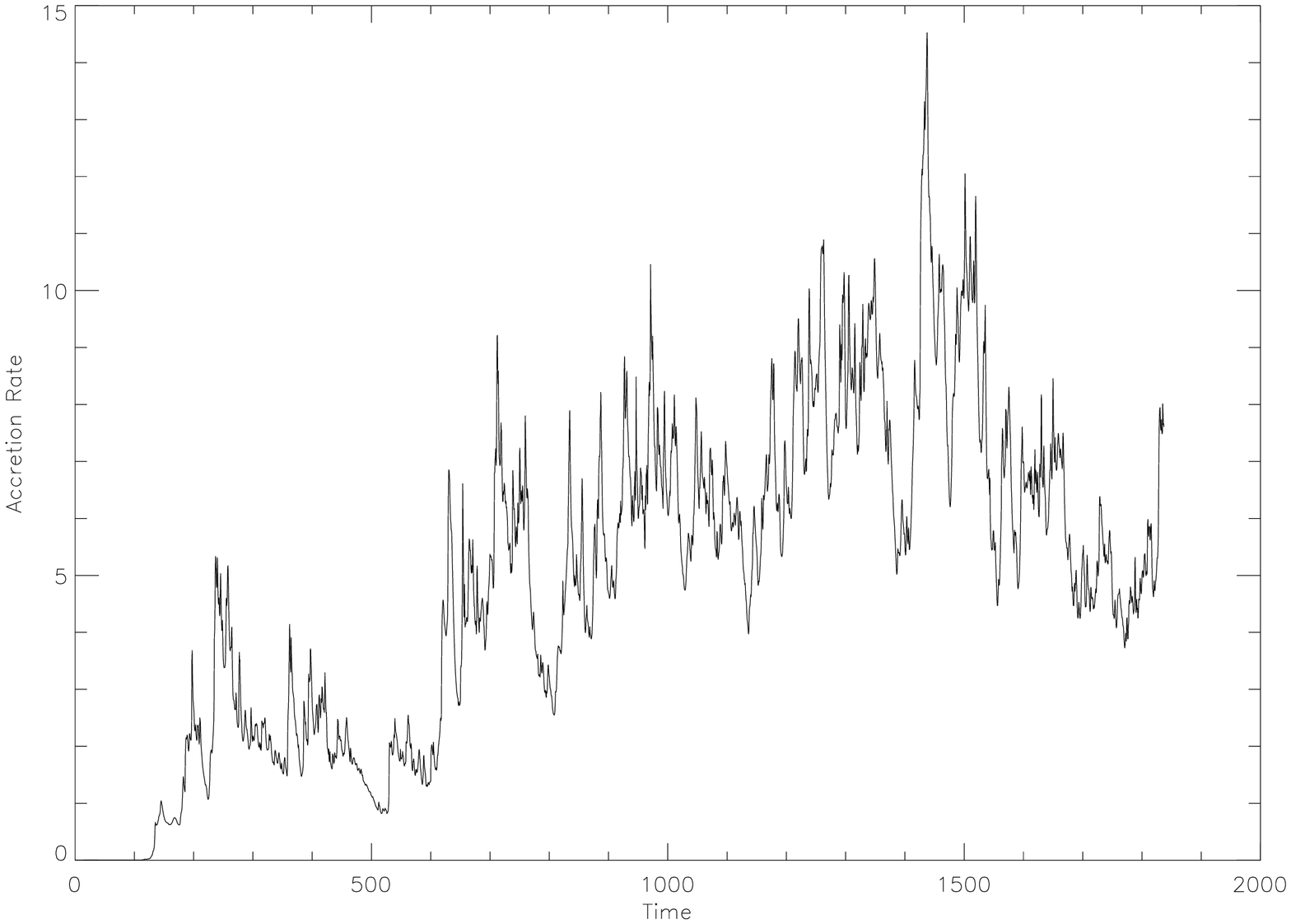,width=5.5in,angle=90}}
\centerline{\psfig{file=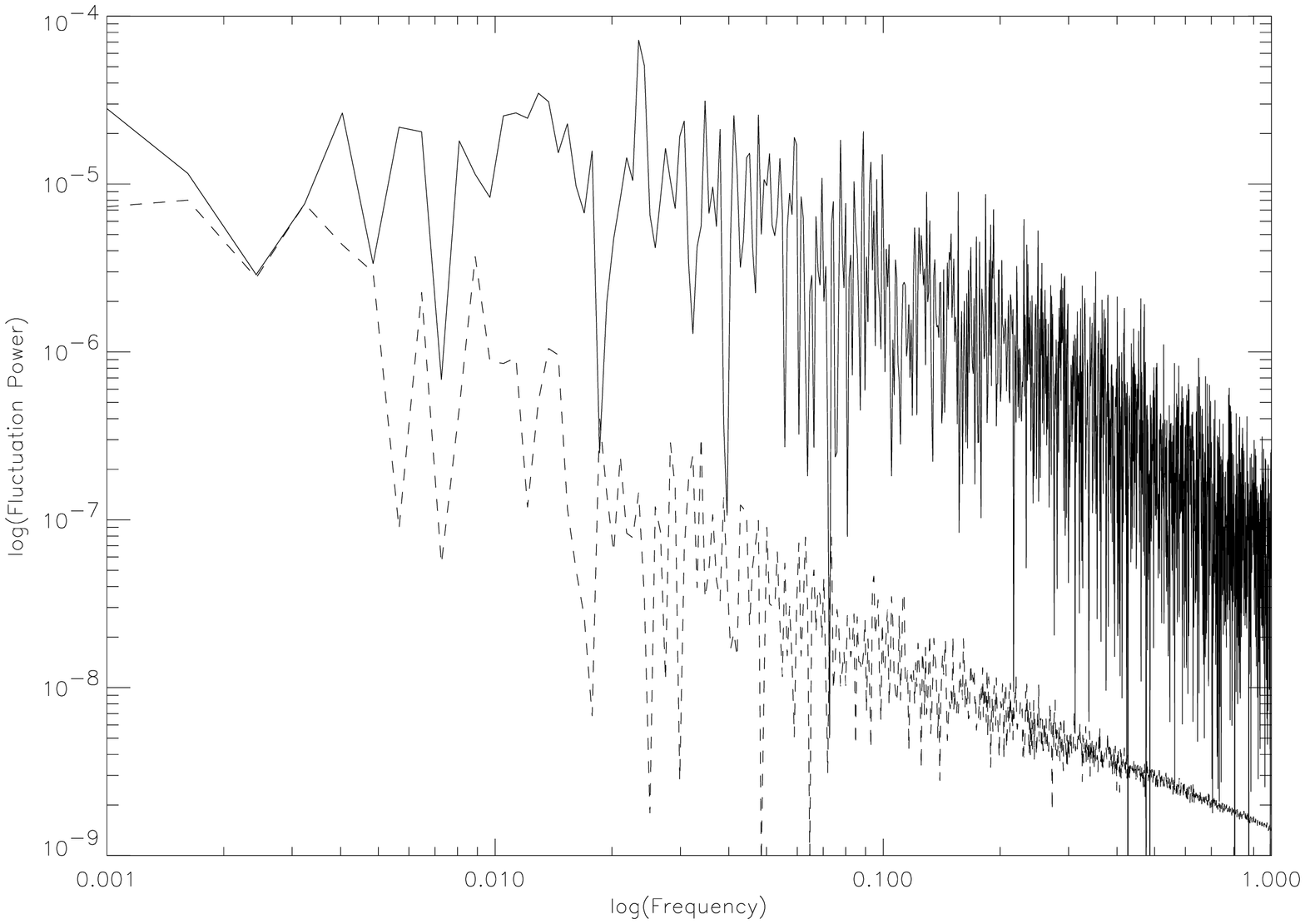,width=5.5in,angle=90}}
\caption{Upper panel: Mass accretion rate at the inner edge as a
function of time in the initially poloidal simulation.  Lower panel:
Solid line is Fourier power density per logarithmic frequency interval
of the accretion rate into the black hole (i.e. $\log(f|\widehat{\dot M}|^2$).
Dashed line is Fourier power in the same units for the volume-integrated
Maxwell stress.
To avoid transients associated with the initial start up and linear growth
phase, the spectrum is computed for $t \ge 600$ time units.
\label{accretehist_p}}
\end{figure}

     We begin quantitative consideration of this simulation with its
mass accretion history.  In the GT4 simulation the mean time-averaged
accretion rate was 4 in code units.  In the HK01 simulation, the mean
accretion rate was very nearly 5; in the new simulation, the mean
accretion rate rises to $\simeq 6.4$ (fig.~\ref{accretehist_p}a).   In
all cases the accretion rate is characterized by large fluctuations with
time.  However, the amplitude of the fluctuations is larger in
the new simulation.  In the previous, lower-resolution work, the
typical peak/trough ratio was $\simeq 1.1$--1.2; in the new one this
ratio approaches 2.

As before, we also examine the Fourier power spectrum of the
time-varying accretion rate.  We previously found that the power
spectrum was a smooth function of frequency that gradually steepened
towards higher frequencies; its mean logarithmic slope over two decades
in frequency was $\simeq -1.7$.  In this new simulation, the power
spectrum is quite similar, but exhibits slightly less curvature.  From
the lowest frequency ($f \simeq 10^{-3}$ inverse time units) up to
roughly the orbital frequency at $r_{ms}$ (0.046 inverse time units)
$|\hat{\dot M}|^2 \propto f^{-1}$; from that frequency up to $f \simeq
1$, the logarithmic slope is $\simeq -2$ (fig.~\ref{accretehist_p}b).
A small local peak appears at around a period of 41, which corresponds
to the circular orbit frequency at $R=4.2$.  The significance of
this peak is unclear; if it is a real feature, it might be interpreted
as suggesting that the final ``plunge" into the black
hole is controlled by events a short way outside the marginally stable orbit.
This would not be surprising as the size of the fluctuations in the MHD
turbulence in this region is of the same order as the local scale
height and the turbulence timescale is comparable to the infall time.
However, it is also possible that the peak is a statistical fluctuation.

   The shape of the Maxwell stress power spectrum is different.  It, too,
may be described as a broken power-law, but in this case
$|\widehat{M_{r\phi}}|^2 \propto f^{-2}$ at high frequencies, while bending
to a steeper slope at low frequencies.  In shearing-box simulations,
the power spectrum of the fluctuating Maxwell stress is also a
power-law of index -2 over several decades in frequency around the
local orbital frequency; it therefore appears that the dominant
stress fluctuations in these global simulations are controlled primarily
by local effects.  Note that the {\it observable} fluctuating
quantity, the luminosity, may be more closely related to the stress
than to the accretion rate because it is tied to the dissipation rate.
However, it is further modified by the distribution of photon escape times.

\begin{figure}[htb]
\centerline{
\psfig{file=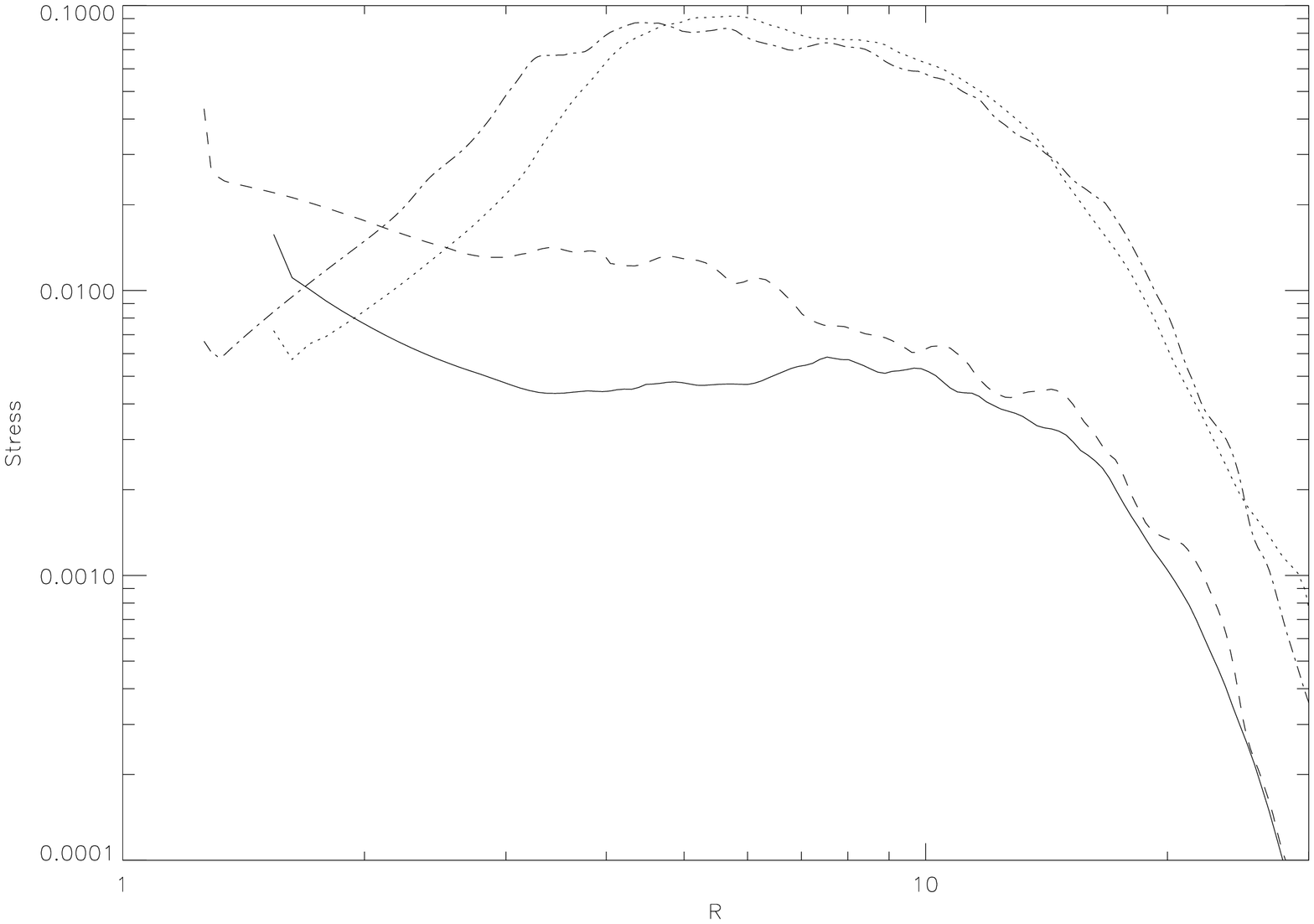,width=6.0in,angle=90}}
\caption{Vertically-integrated and azimuthally-averaged pressure and
magnetic stress at time $t=1490$ in two initially-poloidal
simulations.  The solid curve shows the $R$-$\phi$ magnetic stress in
the simulation of HK01; the dotted curve shows the pressure in the same
simulation.  The dashed and dot-dashed curves show the magnetic stress
and pressure, respectively, in the new higher-resolution simulation.
Analogous data for the initially-toroidal simulation are shown in
fig.~\ref{comparestress_t}.
\label{comparestress_p}}
\end{figure}

     The origin of the larger accretion rate seen in this simulation is
a stronger magnetic field (fig.~\ref{magpres_p}) which produces a
greater $R$-$\phi$ stress.  As we also found in HK01, the field strength
is typically somewhat greater near the disk surface than in the equatorial
plane.  Compared to the HK01 simulation, the
azimuthally-averaged energy density in magnetic field increased by 50\%
in the accreting portion of the disk ($R \leq 10$, $|z| \le 4$); at larger
radius the net flow is outward, and there is little mass or field
at higher altitude.  The stress is also larger than in HK01,
both in absolute terms and as a fraction of the disk
pressure.   Figure~\ref{comparestress_p} compares the
azimuthally-averaged, vertically-integrated magnetic stress and gas
pressure at the endtime of the HK01 simulation to the same quantities
in this new simulation.  In both cases, although the pressure has a
clear maximum as a function of radius (as happens in almost every
disk model due to the sharp inward drop in density as the radial
velocity grows near $r_{ms}$), the
magnetic stress increases monotonically inward.  However, the magnitude
of the stress is everywhere greater in the higher resolution simulation;
it is 2.8 times larger at $r_{ms}$.  In addition, the pressure gradient
is slightly shallower when better resolution is employed.

\begin{figure}[htb]
%\centerline{
%\psfig{file=fig2.ps,width=6.0in,angle=90}}
\caption{The azimuthally-averaged magnetic pressure, $\langle B^2/8\pi
\rangle_\phi$, measured at the end-time in the initially poloidal
simulation, plotted on a logarithmic scale from $10^{-3.5}$ to $10^{-1.5}$.
Fig.~\ref{magpres_t} shows the same quantity in the initially-toroidal
simulation.  \label{magpres_p}}
\end{figure}

   A popular way of parameterizing the importance of the magnetic stress
is through the Shakura-Sunyaev ``$\alpha$" parameter.  In
its original definition (Shakura \& Sunyaev 1973), the disk is
supposed to be azimuthally symmetric and time-steady, and $\alpha$ is
given by the ratio of the vertically-integrated $R$-$\phi$ component of
the stress tensor to the vertically-integrated disk pressure.  Although
it remains useful to compare observed stress levels to the pressure,
there is no unique way (or most physically significant way) to define
$\alpha$ in non-symmetric MHD turbulence.  One may compute
it as a local quantity at each point in $R$, $\phi$, and $z$ or one may
average it in any of a variety of ways.  Different definitions and
averaging procedures yield quantitatively distinct results.  However,
certain trends do show consistent behavior.

    For example, as already suggested in Figure~\ref{comparestress_p},
there is a general rise toward smaller radii in the importance of
magnetic stresses relative to pressure stresses.  Defining
$\alpha_{SS}$ as the azimuthal average of the ratio between the
vertically-integrated magnetic stress $-B_R B_{\phi}/4\pi$ and the
vertically-integrated pressure, we find (see fig.~\ref{alpha_ss_p})
that $\alpha_{SS}$ is typically $\simeq 0.1$ -- 0.2 in the accretion
portion of the disk that is well outside $r_{ms}$
(i.e., $5 \le R \le 10$).  Inside $R=5$, it rises
sharply, reaching $\simeq 0.5$ near $R=3$, and approaching $\sim 10$ at
the innermost edge of the simulation.  In the earlier simulation,
$\alpha_{SS} \simeq 0.06$--0.08 between $R=5$ and $R=10$, rising
inside $R=3$, but always at a lower level than the new simulation.

The ``hump" in $\alpha_{SS}$ around $R=20$ is also seen in the HK01
simulation, but with smaller amplitude.  The hump arises because
the magnetic pressure generally has a larger scale
height than does the gas pressure.  As a result, in the outer portion
of the disk the stress also falls less rapidly with $R$ than
the gas pressure, creating a peak in $\alpha$ outside the gas
pressure maximum.

If one chooses to parameterize the stress by the total pressure, gas
plus magnetic, the rise is less dramatic.  The definition of
$\alpha$ that yields the most constant value is stress divided by
magnetic pressure, $\alpha_{mag} \equiv -2B_R B_\phi/B^2$.  This has a
value between 0.2 and 0.3 throughout the main part of the disk, and
rises slowly toward 1 inside $R=6$.  This definition of $\alpha$
measures the degree of correlation between the toroidal and radial
components of the field.  Because the shear has a consistent sense, the
turbulence is anisotropic and $\langle B_R B_\phi\rangle \neq 0$.  The
value within the disk is typical for stress arising from MHD
turbulence.  However, $\alpha_{mag}$ increases through the plunging
region as turbulence gives way to more coherent fluid flow.  Here
the field evolves primarily through flux-freezing and the
strong correlation created by shear is only weakly diminished by turbulence.

\begin{figure}
%\centerline{
%\psfig{file=fig4.ps,width=6.0in,angle=90}}
\caption{Azimuthally- and time-averaged ratio of the
vertically-integrated stress to: the vertically-integrated gas pressure, i.e.,
$\alpha_{SS}$ (top solid curve); the gas plus magnetic pressure (bottom solid 
curve);
and the magnetic pressure, i.e., $\alpha_{mag}$ (dashed line) in the
initially poloidal simulation.  The time-average runs from $t=1500$
to the end of the simulation.
\label{alpha_ss_p}}
\end{figure}

\begin{figure}
\centerline{\psfig{file=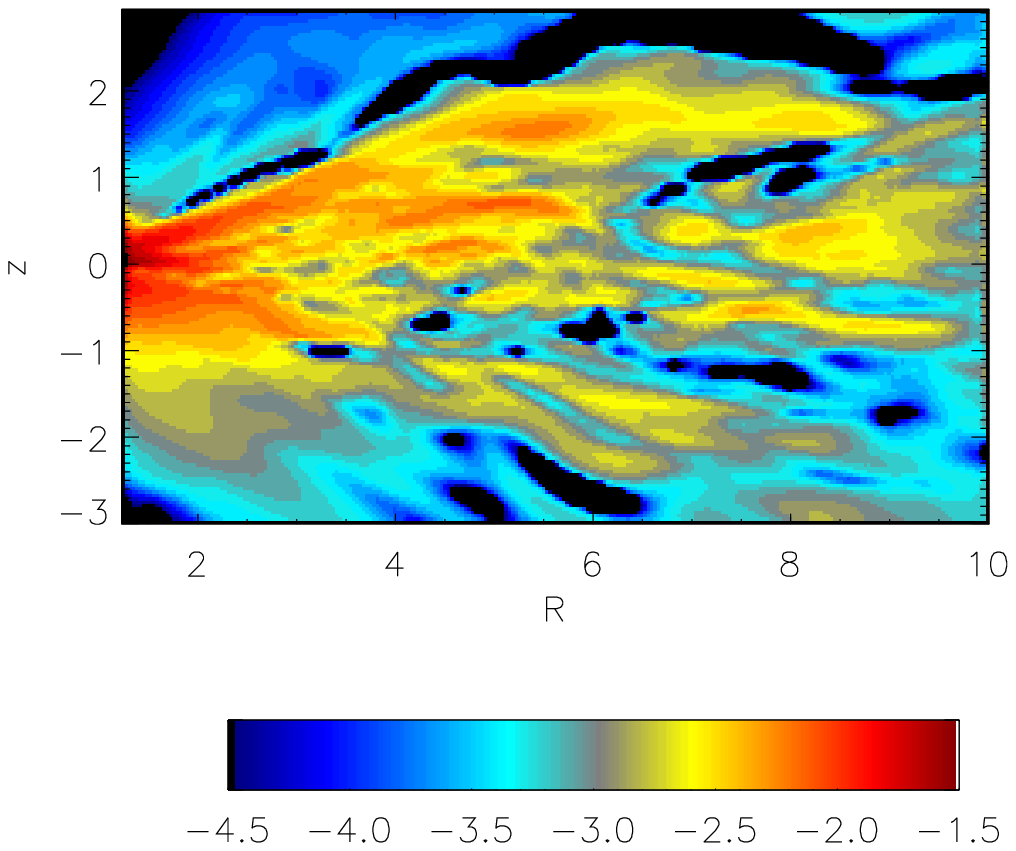,width=6.5in}}
\caption{Logarithm (base 10) of the azimuthally-averaged magnetic
stress at a late-time in the
initially poloidal simulation.  The largest, most consistent stresses
are associated with the plunging region near and within the marginally stable
orbit.  In the black regions of the plot, the azimuthally-averaged
stress is {\it negative}, i.e., it is attempting to transport angular momentum
{\it inward}.  However, the magnitude of the largest azimuthally-averaged
negative stress is at most only $\sim 10^{-3}$ in these units.
\label{stressmap_p}}
\end{figure}

These results highlight the limitations of the $\alpha$ picture:  the
stress doesn't actually correlate particularly well with the pressure.
Another drawback to parametrizing the stress as an averaged $\alpha$ is
that doing so obscures its highly variable nature.  There is tremendous
irregularity in the distribution of stress with altitude and azimuth
(fig.~\ref{stressmap_p}).  At a fixed radius within the accreting
portion of the disk, the azimuthally-averaged stress can vary by more
than an order of magnitude within plus or minus two gas scale-heights
from the midplane; similarly, the vertically-integrated stress may vary
by comparable amounts as a function of azimuth.  Often, but not always,
the stress is greater near the disk surface than in its midplane.

Another way to look at the radial stress distribution is 
with the ``scaled stress.'' In a steady-state disk, the local
vertically-integrated $R$-$\phi$ stress must be equal to the mass
accretion rate times the difference between the local specific angular
momentum $l(R)$
and the specific angular momentum carried into the black hole
$l(r_{g})$, i.e.,
\begin{equation}\label{steadystress}
\bar S = {\dot M \Omega(R) \over 2\pi} \left[1 - {l(r_g) \over l(R)}\right],
\end{equation}
where $\dot M$ is the total mass accretion rate, $\Omega(R)$ the
rotational frequency, and $l$ is the specific angular momentum.
the quantity in square brackets is sometimes called the ``reduction
factor".
 
In HK01, we found that the time- and azimuthally-averaged magnetic
stress very nearly matched the stress predicted by (\ref{steadystress})
for the time-averaged mass accretion rate in the
body of the disk, but, in sharp contrast to the prediction of the
conventional zero-stress model, maintained a high level 
from $R = 5$ to the inner edge of the simulation at $R =
1.5$.  

\begin{figure}
\centerline{\psfig{file=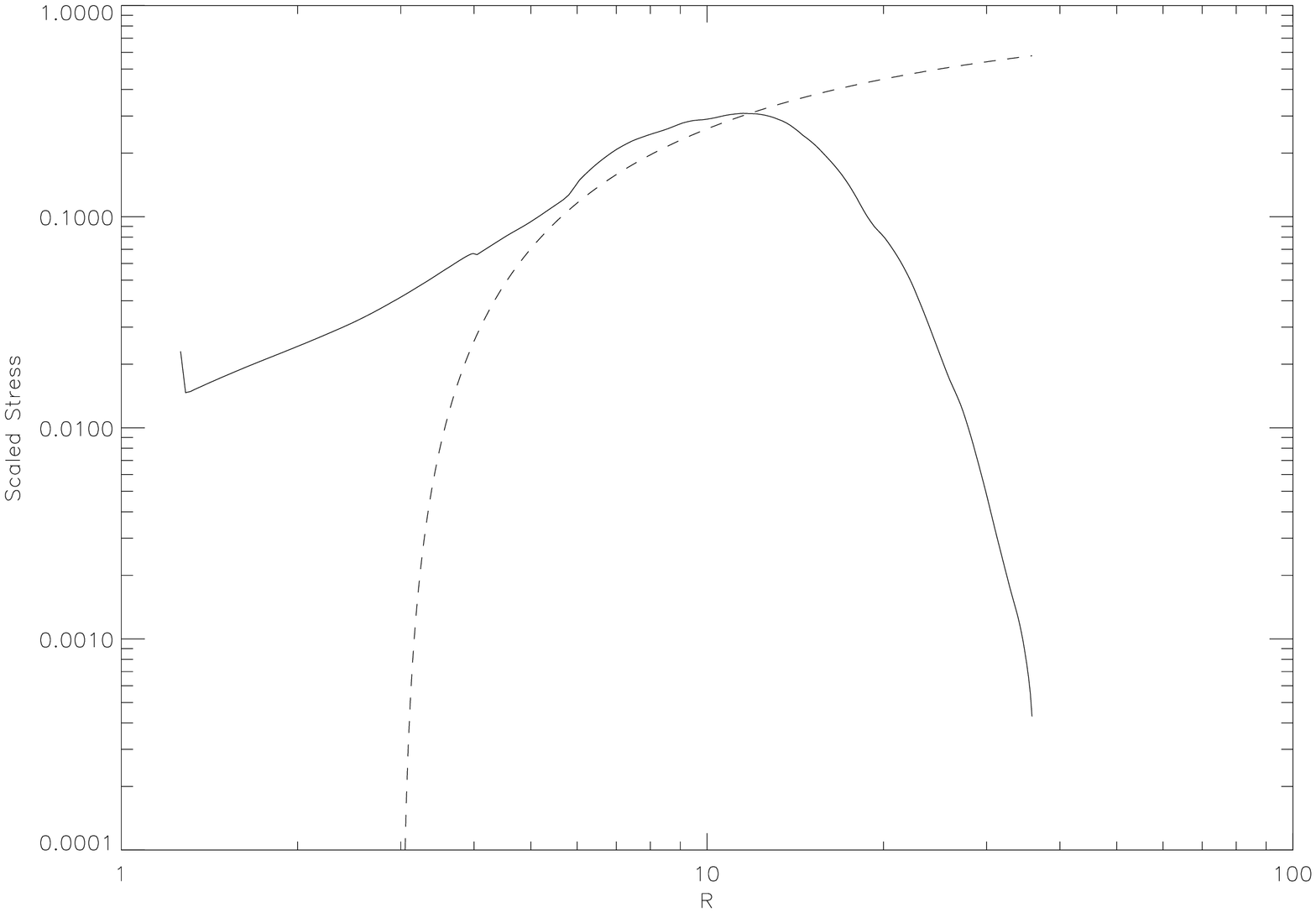,width=6.0in,angle=90}}
\caption{Azimuthal- and time-averaged scaled magnetic stress as a function
of radius (solid curve) contrasted with the prediction of the zero-stress
boundary condition model (dashed curve).  As in Figure~\ref{alpha_ss_p}, the
time average starts at $t=1500$.\label{scaledstress_p}}
\end{figure}

In the higher-resolution simulation the stress behaves in
a similar, but not quite identical, fashion.
In particular, the finer resolution 
leads to increased stress near the marginally stable orbit.
In figure~\ref{scaledstress_p},  the time-averaged,
azimuthally-averaged, and vertically-integrated stress is scaled by
$\langle \dot M \rangle\Omega/2\pi$ in order to highlight its effective
``reduction factor", i.e. the amount by which the stress is diminished
due to the outward angular momentum flux.  Note that for this purpose,
$\Omega$ is the usual orbital frequency for a circular orbit outside
$r_{ms}$, but inside $r_{ms}$ it is the actual orbital frequency as
determined by the angular momentum and radial position of the
material.  In the figure, $\Omega$ inside $r_{ms}$ is approximated by
assuming that the angular momentum is constant in the plunging region;
this approximation (as shown in the following paragraphs) depresses the
result by about 5\% in the mean.

     The scaled stress represents the contrast between the local
specific angular momentum and the accreted angular momentum per
accreted mass.  In a time-steady disk this quantity would approach
unity at large radius, but in the simulation it
is considerably smaller than 1 there because the finite outer edge in our
mass distribution leads to a strong inconsistency there with the
time-steady approximation.  Angular momentum is being transferred into
the outer disk, which is moving outward in response.  By definition,
in a time-steady disk the
scaled stress must be zero at the innermost boundary; the conventional
model predicts that it goes to zero at $R=3$ and stays zero at all
smaller radii.  In the simulation, the scaled stress tracks the
conventional model fairly well between $R=12$ and $R=5$, but at smaller
radii, instead of going sharply to zero at $R=3$, it falls more gradually:
between $R=1.25$ and $R=5$ it is $\propto R^{1.3}$.  This continuing
importance of magnetic stress in
the inner portions of the disk is a counterpart to the increase in
$\alpha_{SS}$ that occurs at the same place.

\begin{figure}
\centerline{\psfig{file=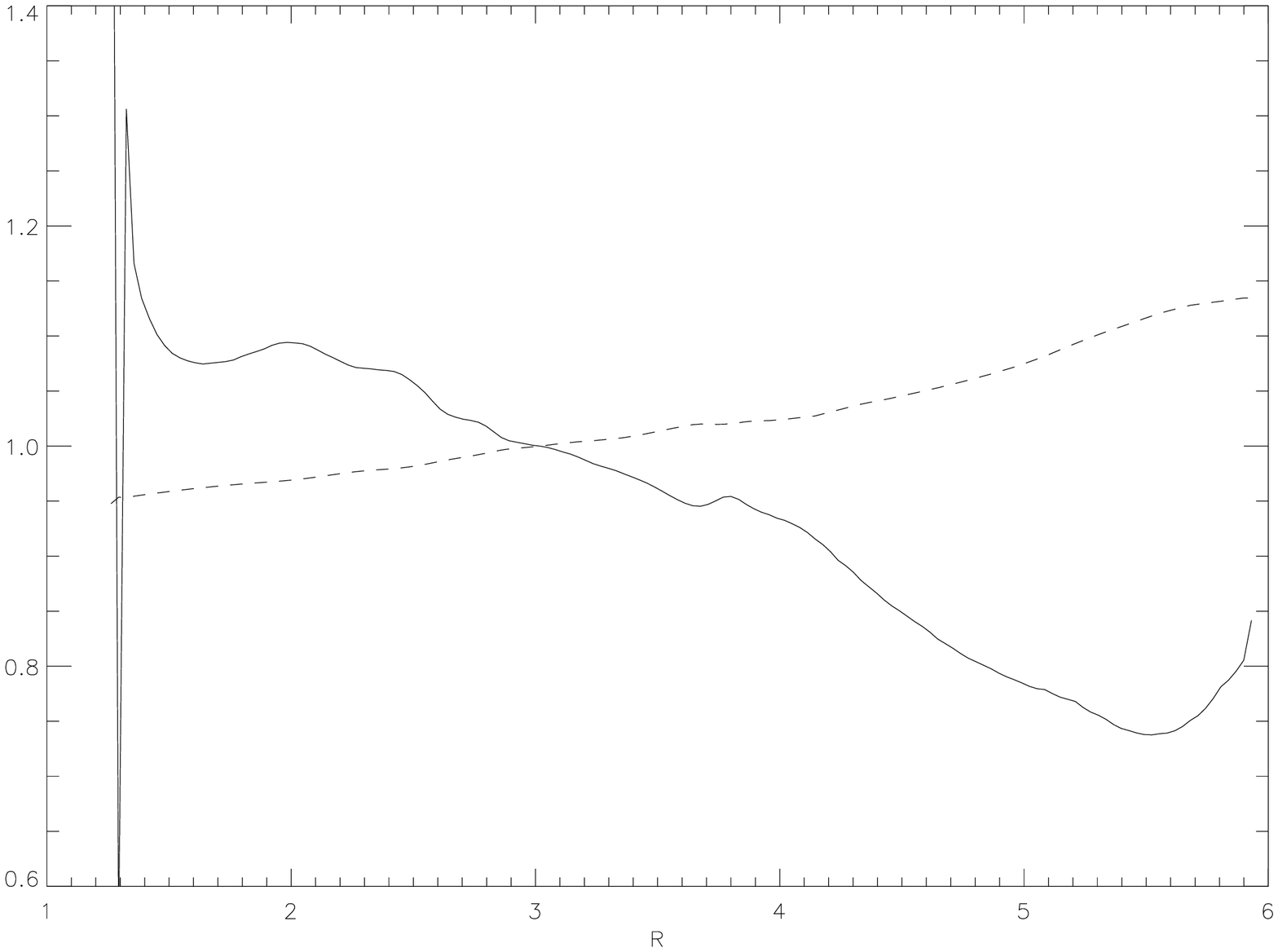,width=6.0in,angle=90}}
\caption{Mass flux weighted azimuthally- and vertically-averaged specific
binding energy (solid curve) and angular momentum (dashed curve) at the end
of the initially poloidal simulation.  Both curves are normalized to their
values at $R=3$ to emphasize the continuing change at smaller radii.
The spikes in the binding energy just outside the inner radius of the
simulation are artifacts.\label{netchange_p}}
\end{figure}

    The result of this continuing stress is a transfer of both angular
momentum and energy from the plunging region inside $R=3$ to the
disk proper at greater radius.  As shown in Figure~\ref{netchange_p},
the mean specific angular momentum falls by about 5\% inside the
marginally stable orbit, while the mean binding energy per unit mass
rises by about 10--20\%.  These significant, but modest, changes in the
mean hide the much larger amounts of angular momentum and energy
transfer that can occur in individual fluid elements.  Some individual
fluid elements arrive at $R=1.3$ with binding energy almost twice the
mean binding energy at $R=3$, while others pass the ``event horizon"
with slightly {\it positive} net energy (fig.~\ref{nete_p}).  Because
the magnetic stress has a vertical scale-height roughly twice the gas
density scale-height, the torque per unit mass felt by matter near the
disk surface is greater than that expressed in the midplane; in
consequence, the specific angular momentum is systematically smaller
off the equatorial plane, often by 10--20\%.
In nonstratified cylindrical disk simulations (Armitage et
al.~2001; Hawley 2001) the decline of $l$ inside of $r_{ms}$ can be
significantly reduced.  A systematic study of the influences of gas
pressure (i.e., $H/R$) and computational domain size carried out by
Hawley (2001) suggests that stratification plays a dominant role in
determining $dl/dR$ inside of $r_{ms}$ and the present simulation
supports that conclusion.
Even with nonstratified disks, however, strong
fields near $r_{ms}$ can produce substantial torques and reductions in $l$
(Reynolds, Armitage \& Chiang 2001).

\begin{figure}
%\centerline{\psfig{file=fig8.ps,width=5.0in}}
\caption{Net energy in rest-mass units in a slice through the
equatorial plane in the plunging region at late time in the initial
poloidal field simulation.  Note that in these units the binding energy
of a circular orbit at $r_{ms}$ is 0.0625.
\label{nete_p}}
\end{figure}

Figure \ref{nete_p} illustrates another aspect of the character of the flow
inside $r_{ms}$: as gas plunges toward the ``event horizon", transfers of
energy and angular momentum between adjacent fluid elements become
stronger and stronger.  Orbital shear stretches initially coherent
regions into spirals stretching roughly a radian in azimuth.  Local
contrasts become especially strong deep in the plunging region
because gas there can no longer exert forces and torques on
gas farther out.

\begin{figure}
\centerline{\psfig{file=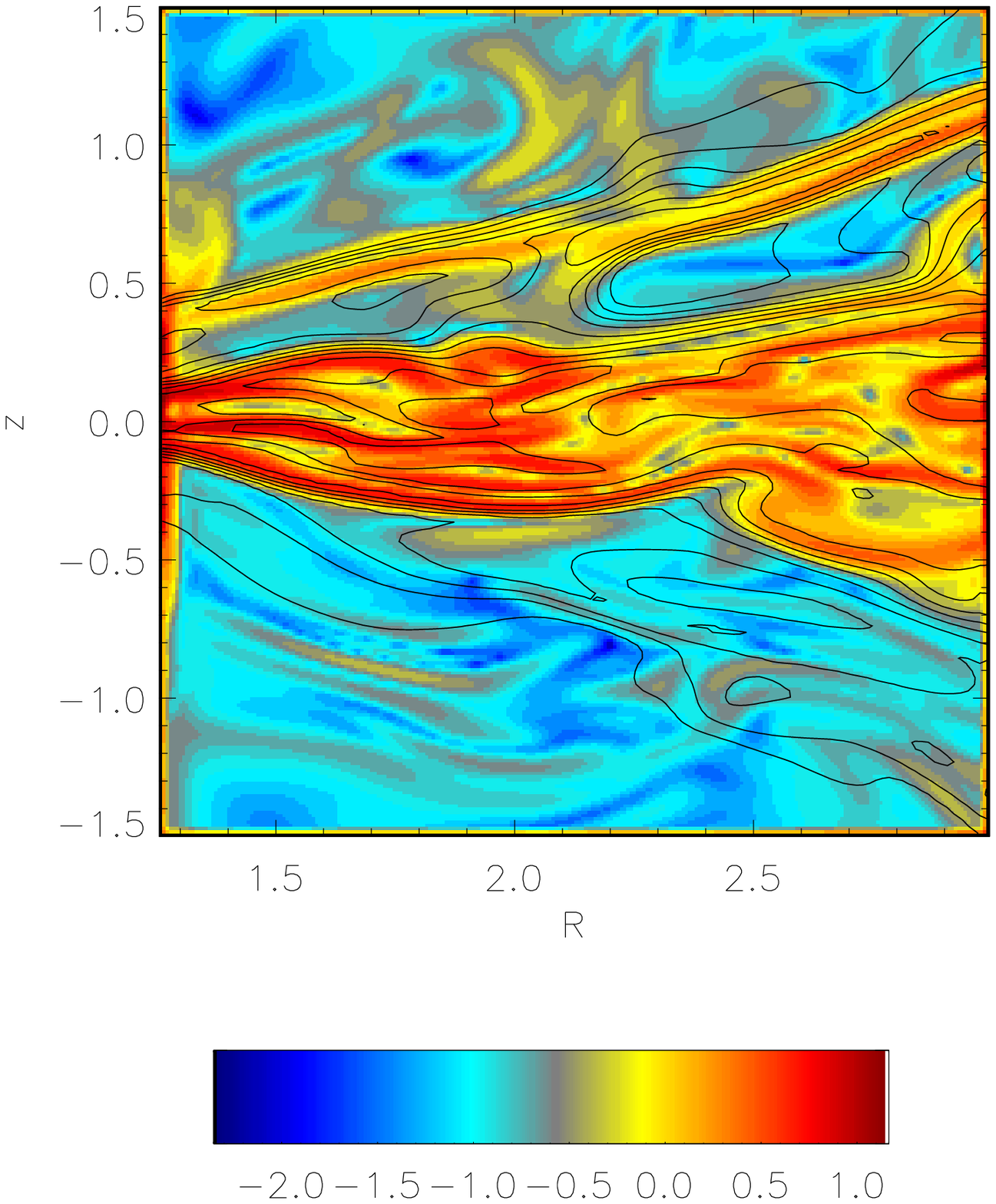,width=5.0in}}
\caption{Magnitude of the electric current (color contours) compared
with gas density (line contours) in a poloidal slice at $\phi = \pi/4$,
both in a logarithmic scale.  The two quantities are very well
correlated.
\label{mag_v_dens}}
\end{figure}

Viewed in a poloidal slice (fig.~\ref{mag_v_dens}), the simulation
reveals a different effect: in the plunging region, gas is concentrated
where the magnetic shear or current density is particularly great.
Interestingly, this concentration is much less noticeable considered as
a function of azimuth.  Generally speaking the gas and
magnetic pressures are anticorrelated in the disk (although within the
disk the gas pressure is generally larger).  The total pressure is much
smoother than either the gas or magnetic pressure separately.

\section{Initially Toroidal Field}

   The second simulation begins with the same initial torus as before,
but with a purely toroidal magnetic field.  The most appropriate field
topology in accretion disks is not known {\it a priori} and this
simulation investigates how much the results depend on the initial
field structure.  We believe (as we discuss at greater length below)
initially toroidal and initially poloidal fields bracket
the range of possibilities.

    The initial field configuration has toroidal field with a
plasma $\beta=10$ wherever $\rho \ge 0.1$.  The model was run at two
resolutions.  One was the same as described above with $256\times
64\times 192$ zones.  This disk was evolved for 800,000 timesteps to
time $t=2704$ (124 orbits at $r_{ms}$).  The other, a lower resolution
comparison simulation, is discussed below in \S5.1.

The toroidal field simulation evolves in significantly different ways
from the initially poloidal field case.  First, with no initial radial
field there is no shear amplification of the toroidal component during
the early stages of the evolution.  Second, the most unstable
wavenumbers of the linear toroidal field MRI are different from those
of the vertical field.  For a pure toroidal field the fastest growth
rates correspond to azimuthal wavenumber $m \sim v_{\phi}/v_A$ (i.e.,
$m$ is large for typical subthermal field strengths) and large poloidal
wavenumbers (Balbus \& Hawley 1992; Terquem \& Papaloizou 1996; Kim \&
Ostriker 2000).  In previous local and global simulations done with a
toroidal field (Hawley, Gammie, \& Balbus 1995; H00), small scale
fluctuations appear first, followed gradually by larger scale
turbulence.  As a result of these effects, the torus evolves more
slowly than with a poloidal field, and the field energy and stress
levels at saturation are smaller.  In the present simulation the period
of linear growth and evolution occupies the first $\sim 1000$ time
units.

Nonetheless, the evolution of the disk after $\sim 1000$ units is in
many ways qualitatively similar to the initially-poloidal simulation.
Angular momentum is transported outward and its distribution
evolves toward Keplerian.  Smaller field energies and stresses cause
the toroidal disk to be thinner than the poloidal field disk, with
$H/R$ remaining comparable to the initial value 0.12 at $R=10$ and $H/R
\simeq 0.1$ inside $R=5$.  Just as for the initially-poloidal
simulation, there is substantial accretion through the inner
cylindrical boundary and hardly any mass loss through the outer
boundaries.

\begin{figure}[htb]
\centerline{
\psfig{file=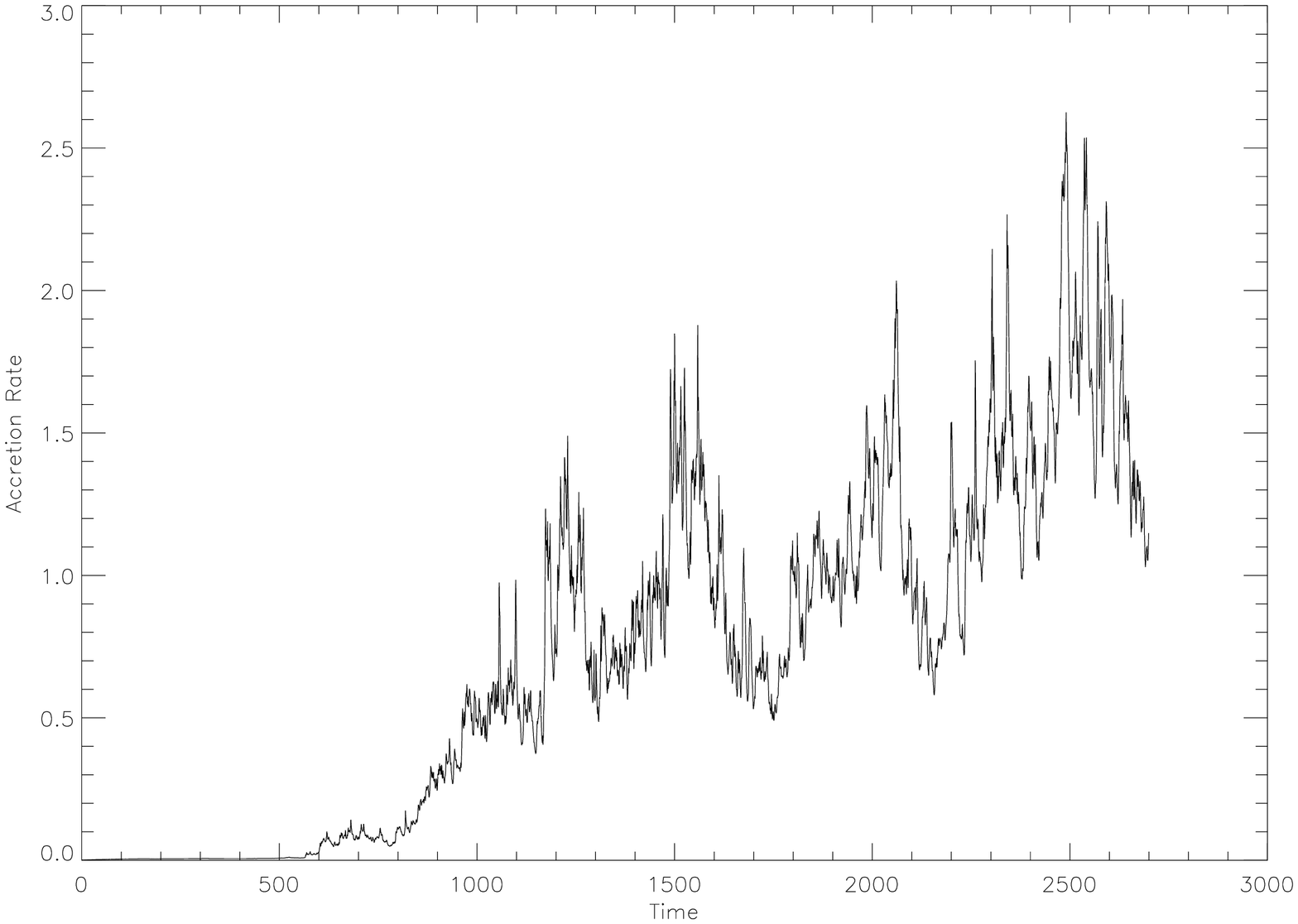,width=5.0in,angle=90}}
\centerline{\psfig{file=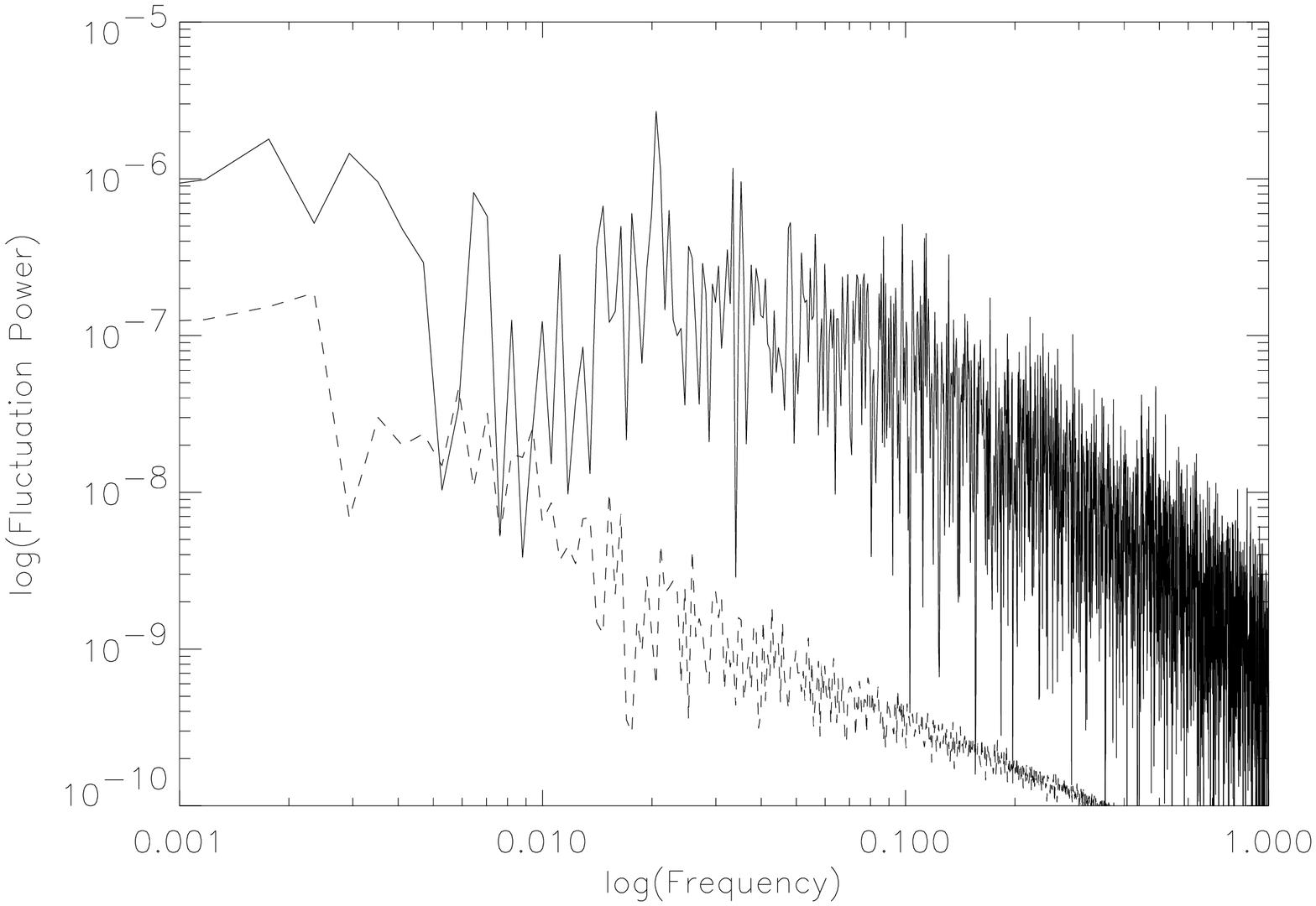,width=5.5in,angle=90}}
\caption{Upper panel: Mass accretion rate at the inner edge as a function of
time in the initially toroidal simulation.  Lower panel: Fourier power density
per logarithmic frequency interval of the accretion rate into the black hole
(solid curve), and Fourier power density per logarithmic frequency interval
of the volume-integrated Maxwell stress (dashed curve). To
avoid transients associated with the initial start up and linear growth
phase, the spectrum is computed for $t \ge 1000$ time units.
\label{accretehist_t}}
\end{figure}

    From $\sim 1000$ time units to the end, the accretion rate ranges
between $\simeq 0.5$ and 2.5, i.e., between 8\% and 40\% of the
accretion rate in the initially poloidal simulation
(fig.~\ref{accretehist_t}).  In contrast to the accretion rate history
of the poloidal simulation, $\dot M$ builds slowly and features
relatively smooth swings from a ``low state" to a ``high state" and
back again.  The ratio between the high and low accretion rates can be
as large as $\sim 3$; there is no time at which $\dot M$ could fairly
be said to be approximately stationary.  In the poloidal case there are
also large fluctuations between high and low rates, but these occurred
on shorter timescales.  Here there seems also to be a secular increase
in $\dot M$ after $t=1000$, but its significance is hard to gauge
as any long-term trend is masked by the very large fluctuations that
occur.  Figure \ref{accretehist_t}b is the Fourier power spectrum of
the initially-toroidal accretion rate.  It strongly resembles the power
spectrum of the initially poloidal simulation, both with regard to its
broken power-law character and the existence of a small peak (which in
this simulation is at the slightly greater period of 47, corresponding
to the circular orbit frequency at $R=4.5$).  A second peak can be seen
at a period of 142, which is the circular orbital frequency associated
with $R=8.7$.  The significance of these peaks is as uncertain as that
of the similar peak found in the Fourier power spectrum of the
initially-poloidal accretion rate.

    Because the accretion rate varies so much during this simulation,
no single time can fairly represent its behavior.  We will discuss
two ``snapshots", one from a low accretion rate stage
(at 2704 time units, when the accretion rate through the inner edge was
$\simeq 1$), the other from a time of high accretion rate (2537 time
units, when the accretion rate was 2.4).

In previous initial toroidal field simulations, both global and local,
saturation occurs at lower turbulent field energies than models
beginning with poloidal components.  This is true in the present
simulation as well.  The azimuthally-averaged field strength
(fig.~\ref{magpres_t}) is about 10\% stronger in the ``high rate" case
compared to the low.  Both, however, are about six times smaller than
in the poloidal simulation when averaged over the accreting portion of
the disk ($R \leq 10$, $|z| \leq 4$).  In addition, the vertical scale
height of the magnetic field is roughly half what it was in the
initially poloidal case.

\begin{figure}[htb]
%\centerline{
%\psfig{file=fig11a.ps,width=5.0in,angle=90}}
%\centerline{
%\psfig{file=fig11b.ps,width=5.0in,angle=90}}
\caption{Azimuthally-averaged magnetic energy density in the ``high
accretion rate" snapshot (upper panel) and the ``low rate" snapshot (lower
panel).  The scales in these two figures are identical to the scale used
in fig.~\ref{magpres_p} to ease comparison.
\label{magpres_t}}
\end{figure}

    Not surprisingly, given the generally weaker magnetic field, the
effective $\alpha_{SS}$ is significantly smaller in the toroidal case
than in the poloidal.  In the poloidal simulation
(fig.~\ref{alpha_ss_p}), $\alpha_{SS} \simeq 0.1$ at $R=10$ and rises
sharply inward to a peak of almost 10 at the inner boundary.  In the
two toroidal snapshots, $\alpha_{SS} \simeq 0.01$ -- 0.03 near $R=10$,
and, although rising inward, it reaches a maximum of only $\simeq
0.2$--0.3 just inside the marginally stable orbit.  This contrast, of
course, accounts for the lower accretion rate in the toroidal case,
despite having initially an identical mass surface density.  The value
of $\alpha_{mag}$, the ratio of the magnetic stress to the magnetic
pressure, is 0.2--0.3 in the disk, rising from that value to $\sim 1$
from $R=4$ to the event horizon.  This is similar to the poloidal
field case, except that the systematic increase begins at a slightly
smaller radius.

\begin{figure}[htb]
\centerline{
\psfig{file=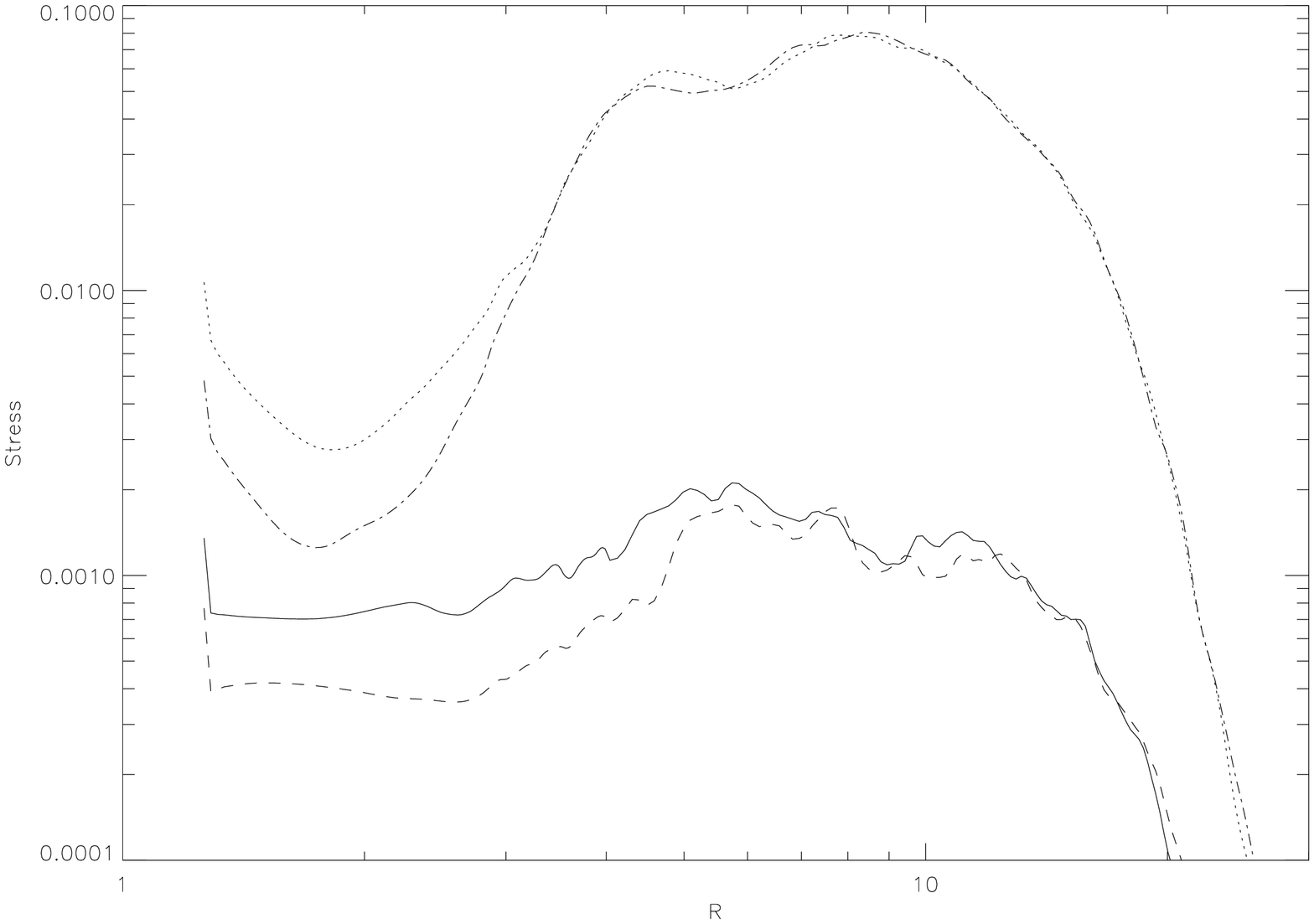,width=6.0in,angle=90}}
\caption{Vertically-integrated and azimuthally-averaged pressure and magnetic
stress at late times in the high and low accretion rate snapshots from
the initially toroidal simulation.  The solid curve shows the
$R$-$\phi$ magnetic stress in the high accretion rate case; the dotted
curve shows the pressure at the same time.  The dashed and dot-dashed
curves show the magnetic stress and pressure, respectively, in the low
accretion rate snapshot.\label{comparestress_t}}
\end{figure}

     Comparatively weaker magnetic field also leads to a different
shape to the stress distributions (fig.~\ref{comparestress_t}).
The radial pressure gradient is much larger in the toroidal field case
than in the poloidal field simulation: the vertically-integrated
pressure falls by roughly a factor of 30 from $R=10$ to $R=3$, whereas
the decline was only about a factor of 3 in the initially poloidal
run.  The magnetic stress, which rose steadily inward in the poloidal
case, is approximately flat here between $R=15$ and $R=5$ and falls by
about a factor of two from $R=5$ to $R \simeq 2$--3, inside of which
it is again constant.

\begin{figure}
\centerline{\psfig{file=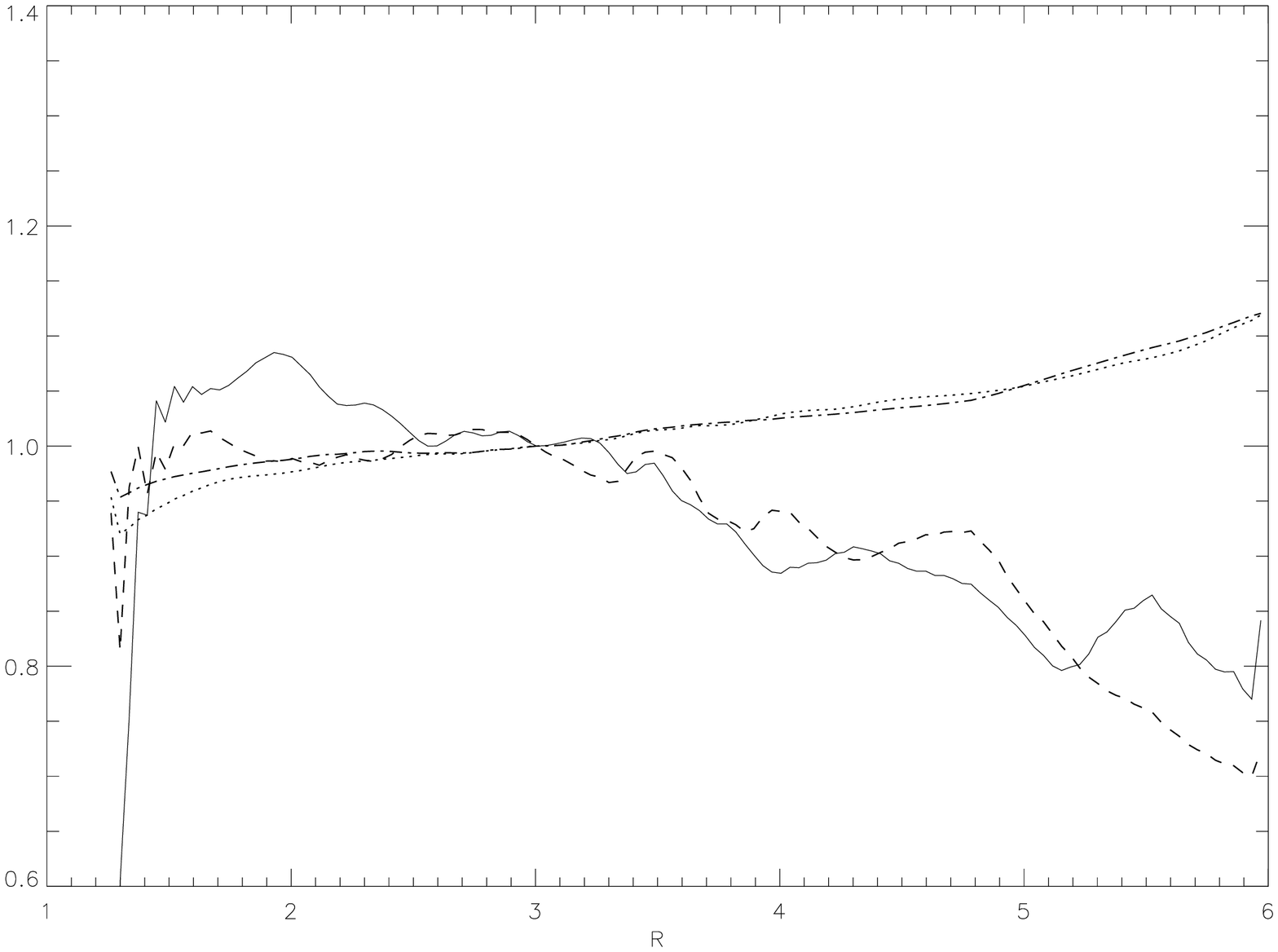,width=6.0in,angle=90}}
\caption{Mass flux weighted azimuthally- and vertically-averaged specific
binding energy and angular momentum as functions of radius in the
initially toroidal simulation.  The solid curve shows binding energy
in the high accretion rate state, the dotted curve angular momentum;
the dashed and dash-dotted curves are energy and angular momentum,
respectively, in the low accretion rate state.  All four curves are
normalized to their
values at $R=3$ to emphasize the continuing change at smaller radii.
The spikes in the binding energy just outside the inner radius of the
simulation are artifacts.\label{netchange_t}}
\end{figure}

   Although the magnetic stress is generally weaker in the toroidal
simulation than in the poloidal one, the mean change in specific
angular momentum inside the marginally stable orbit is similar to that
found in the poloidal simulation: a drop of 5 -- 10\%
(fig.~\ref{netchange_t}).  On the other hand, the detailed character of
the angular momentum distribution in the plunging region is quite
different in the initially toroidal simulation.  As we have already
remarked, the change in energy and angular momentum of individual fluid
elements is much greater than the mean change.  When the accretion rate
is especially high in the toroidal simulation, the contrast in specific
angular momentum between adjacent fluid elements is much greater than
when the accretion rate is low (fig.~\ref{angmomcontrast}).  Instead of
passing through the inner boundary with specific angular momentum 2.6
(the angular momentum of the marginally stable orbit), in the high
accretion rate case there are streams arriving with as little as
$\simeq 1.8$ and some that arrive with as much as $\simeq 3$.  By
contrast, during the time of low accretion rate, the range is only from
$\simeq 2.2$ -- 2.6.

    The energy of accreting matter behaves in similar fashion, but with
some notable contrasts.  At the time of high accretion rate, the mean
binding energy increases gradually from $R=3$ to $R=2$ to a maximum
that is about 10\% greater than at the marginally stable orbit, while
at the time of low accretion rate there is little change in mean
binding energy in the plunging region.  However, just as in the
poloidal simulation, the slow change in mean energy masks very large
changes in the energy of individual fluid elements
(fig.~\ref{nete_t}).  As with the angular momentum, the energy exchange
inside $R=2$ is much stronger in the case of initially toroidal field:
whereas the maximum increase in binding energy in the poloidal case was
about a factor of two, in the toroidal case it is a factor of 10 --
20!  Finally, in all cases, there is a spike in the mean binding energy
just outside $R_{min}$ that is an artifact of the boundary condition.

\begin{figure}
%\centerline{\psfig{file=fig14a.ps,width=3.0in}}
%\centerline{\psfig{file=fig14b.ps,width=3.0in}}
\caption{Mass-weighted vertically-averaged specific angular momentum
in the inner part of the accretion flow for the initially toroidal
case: upper panel, at high accretion rate; lower panel, at low
accretion rate.\label{angmomcontrast}}
\end{figure}

\begin{figure}
%\centerline{\psfig{file=fig15.ps,width=5.0in}}
\caption{Net energy in rest-mass units in a slice through the equatorial
plane in the plunging region during the time of high accretion rate in the
initially toroidal simulation.
\label{nete_t}}
\end{figure}

\section{Discussion}

\subsection{Field topology}

\subsubsection{Stress distribution}

    We suggest that the contrasting behavior seen in the toroidal
and poloidal field simulations illustrates the effects of different
inflow and magnetic dissipation times.  The ratio between these
two quantities distinguishes the disk proper from the plunging
region.  In the disk proper, the inflow time is long compared to
the dissipation time so that turbulent dissipation controls the
magnetic stress; on the other hand, in the plunging region the inflow
time is short compared to the dissipation time so that the field and
stress evolve primarily through flux-freezing.

     The shearing-box approximation corresponds to the disk proper because
the inflow time is effectively infinite.  In those simulations (Hawley,
Gammie \& Balbus 1995, 1996), the magnetic field tends to saturate at
a roughly constant $\beta$ value with respect to the local pressure.\
The stress is then a fixed fraction $\alpha_{mag}$ of the field energy density.
We find just this sort of behavior in the portion of our simulations
well outside $r_{ms}$, with $\alpha_{mag} \simeq 0.2$--0.3 in the
initially-poloidal simulation and somewhat smaller in the initially-toroidal
case.

     However, in our simulations $\alpha_{mag}$ increases steadily
toward unity from $R\simeq 5$ to the inner boundary, indicating a
qualitative change in the nature of the
accretion flow at this radius.  At small radii, the inflow time
diminishes while the dissipation time (we expect) does not change
greatly.  Where the inflow time is shorter than the dissipation
time, the field evolves via shear amplification and flux-freezing and
is decoupled from the local pressure.  The dominance of coherent
shear over turbulence accounts for the increase in $\alpha_{mag}$.

    The contrast between the initially-poloidal and initially-toroidal
cases arises from the fact that the inflow time at
any given radius of the disk proper is longer in the toroidal
simulation than in the poloidal due to the weaker magnetic stress.
As a result, the zone in which the field energy density roughly
tracks the pressure extends further inward when the initial field
is purely toroidal.  The magnetic stress in the
toroidal run begins to evolve via flux-freezing at a smaller radius
than in the poloidal run and (crudely) follows the diminishing pressure
between $R=6$ and $R=3$, whereas in the poloidal run the magnetic
stress continues increasing inward even when the pressure is
decreasing.

In evaluating these contrasts, however, it is
essential to remember that the dissipation rate in these simulations is
controlled by numerical, not physical effects (see further discussion
in \S 5.2); consequently, our results may not be quantitatively
reliable guides to the behavior of real disks.

\subsubsection{Which field topology is ``natural"}

  From these simulations it appears that the level of stress at and
inside the marginally stable orbit depends significantly upon whether
the initial field topology is toroidal or poloidal.  Dependence of the
level of MHD turbulence on initial field topology has been observed in
previous simulations, both global and local.  The linear analysis of
the MRI shows that the fastest growing wavelengths of vertical fields are
axisymmetric and have larger vertical wavelengths, while the toroidal
field instability is nonaxisymmetric, favors small wavelengths, and
grows transiently as time-dependent radial wavenumbers swing from
leading to trailing.  Given these differences it is important to ask
which field topology better represents the state of an actual accretion
flow.

A result that is common to all of the simulations done to date,
both global and local, is that the toroidal field energy dominates the
final turbulent state, whatever the initial field configuration.  This
does not mean, however, that the initial purely toroidal field
evolution is the best model; it is in many ways a singular state, and
there are sharp differences between it and models that have even a small
amount of local net poloidal field.  For example, Hawley et al.\ (1995)
carried out several local shearing box runs in which a very much weaker
vertical field was added to boxes with strong toroidal fields.  The
result was greatly enhanced turbulence.  As noted by Hawley (2000) even
if global simulations begin with no overall net poloidal field, but
include some poloidal field in the form of field loops, local regions
of the disk still behave like local simulations with poloidal fields,
that is they generate strong turbulence and $\alpha \sim 0.1$.

Hence, if an accretion flow begins from a source with even a small amount of
poloidal field, and that field remains in or is amplified by the
resulting inflow, one would expect local regions of the disk to behave
more like the poloidal field simulation than like the pure toroidal
field case.  Given the strong influence of even a weak poloidal field,
the inner disk dynamics could well be variable over long timescales due
to variations in the net poloidal field carried in by accretion from
the outer disk.  Nonetheless, in a time-average sense we expect real disks
to resemble the initially-poloidal simulation more than the
initially-toroidal simulation.

\subsection{Influence of resolution}

The contrast between our new results and those obtained from the
previous simulations of HK01 and H00 provides a measure of the effects
of numerical resolution (Table~\ref{simres}).  The grid used in HK01
is, in its inner parts, finer by a factor 1.9 in $\Delta R$ and 2.5 in
$\Delta z$ relative to the grid of simulation GT4 in H00; our new grid
is finer by a factor 3.3 in $\Delta R$, a factor 2.4 in $\Delta z$ and
2.0 in $\Delta \phi$ relative to the simulation of HK01.  We have also
done a second initially-toroidal simulation with lower resolution than
the one described in \S 4.  It used $128\times 32\times 128$ grid
zones.  The radial grid concentrated 55 zones linearly spaced inside
$R=4$; outside this point $\Delta R$ gradually increased to the outer
boundary at $R=32$.  The vertical grid put 40\% of the zones within
$\pm 2$ of the equator, with a graded mesh beyond that point out to $z=
\pm 8.8$.  Thus in the lower resolution toroidal simulation $\Delta R$
is twice as large, and $\Delta z$ three times as large in the inner disk
region.  This model was run for 400,000 timesteps to time $t=3873$.

\begin{deluxetable}{cccc}
\tablewidth{0pt}
\tablecaption{Inner Disk Region Resolutions\label{simres}}
\tablehead{\colhead{Simulation} & \colhead{$\Delta R$} & \colhead{$\Delta z$}
& \colhead{$\Delta \phi$} }
\startdata
GT4 (H00) & 0.16 & 0.16 & 0.05\\
HK01 & 0.083 & 0.0625 & 0.05\\
new poloidal & 0.025 & 0.026 & 0.025\\
new toroidal & 0.025 & 0.026 & 0.025\\
low-res. toroidal & 0.05 & 0.077 & 0.05\\
\enddata
\end{deluxetable}

It should be noted that, since this is a study of the disk structure
near $r_{ms}$,  we have focused the resolution improvements within the
inner regions of the flow and around the equator.  To the extent that
the overall evolution depends on the outer parts of the disk, the
effective resolution difference between simulations is far less than
that of the inner region.  The resolution in the outer disk of HK01
is actually poorer than that in the same portion of GT4.

\subsubsection{Accretion rate}

    With two successive improvements in resolution by roughly factors
of 3, the mean accretion rate in the initially-poloidal case
increased 60\% from GT4 to HK01 and a further 30\% in the new
simulation.  The increase in accretion rate between HK01 and the new
simulation might have been slightly greater if the new simulation had
spanned a full $2\pi$ in azimuthal angle, rather than being restricted
to only a single quadrant.  Moreover, the fluctuations about the mean
also increased substantially with improving resolution, with the
peak to trough ratio increasing from $\simeq 1.1$ to $\simeq 2$.

    A similar rise in accretion rate with increasing resolution was
seen in the pair of initially-toroidal simulations.  At the lower
resolution, the time-averaged accretion rate was $\simeq 0.7$,
fluctuating between a high of 1.7 and a low of 0.3.  By contrast, with
better resolution the accretion rate (after the end of transient
effects) had a mean of 1.7 and fluctuated between 0.5 and 2.6.

These numerical experiments show that increasing resolution produces
increased stress and accretion rates.  For the poloidal field
simulation the observed time-steady mean accretion rate is probably
within tens of percent of the ``converged'' value, although almost
certainly still on the low side.  However, we have not yet achieved
sufficient resolution to gauge the convergence of the magnitude of the
fluctuations around the mean.  For the simulation with the initial
toroidal field, the increase in $\dot M$ with resolution was
sufficiently great that it is not yet possible to estimate what the 
converged value might be.

\subsubsection{Magnetic field}

   Between GT4 and HK01 there was a dramatic increase in magnetic field
energy density in the inner disk: roughly a factor of 10.  However,
the improvement in resolution between HK01 and the new simulation,
although comparable in magnitude to that between GT4 and HK01, changed
the magnetic field much less: an increase of about 50\% in mean energy
density.

   Analogous effects can be seen in the toroidal case.  Growth to
turbulent saturation takes almost twice as long, and the averaged poloidal
field energies (a measure of the strength of the turbulence) in the saturated
state are reduced by about a factor of 2 in the poorer resolution
simulation.

    Stronger magnetic field is mirrored in a rising ratio between
magnetic stress and gas pressure.  In the stable part of the accretion
region (i.e., $5 \le R \le 10$), $\alpha_{SS} \simeq 0.05$--0.1 in HK01 
and 0.1--0.2 in the new poloidal simulation.
The initially-toroidal simulations show an increase
from $\simeq 0.01$--0.028 in the lower-resolution simulation to 
$\simeq 0.02$--0.03 in the higher-resolution run.

It appears very likely that improvements in resolution reduce numerical
dissipation of magnetic field.  Because of the central role of magnetic
torques in driving accretion, the stronger magnetic field then
increases the accretion rate.   On the basis of the decline in the rate
of increase of the magnetic field energy density with resolution in
going from HK01 to our new poloidal simulation, it appears that, with
regard to the mean field intensity, we are approaching convergence
for the poloidal case, but it is likely that further refinement would yield
at least modest quantitative changes.

  We have less confidence that the initially-toroidal case is
sufficiently well-resolved.  As noted above, the toroidal field MRI
favors high wavenumbers, precisely those that are the most challenging
to resolve.  The resolution comparison carried out here suggests that
even with the higher resolution grid the simulation is not yet
converged, and the values reported here must be regarded as lower
bounds.

\subsubsection{Stress distribution}

   As we discussed at length in \S 5.1, the ratio between effective magnetic
dissipation time and inflow time plays a critical role in governing the
magnitude of stresses in the inner disk.  Where the dissipation time is
short compared to the inflow time, the magnetic field strength is
coupled to the local pressure; where the comparison is reversed, the
field evolves by flux-freezing.  At even our finest resolution,
the dissipation rate is exaggerated
by numerical effects.  Consequently, in these simulations the magnetic
field intensity is tied to the pressure further in 
than would be expected in a real disk where flux-freezing would set in at
a larger radius.  Thus, the stress observed
in these simulations at small radii is most likely still undervalued.

\subsubsection{Energy conservation}

Energy conservation and energy flow can be difficult for a finite
difference simulation to compute accurately when there are significant
departures from equipartition in energy components.  Here the bulk of the
energy is gravitational and orbital; thermal energy is only 2\% of the
total at the beginning of the simulation, and 4\% at the end.  Thermal,
magnetic, and poloidal kinetic energy combined are only 14\% of the
circular orbit binding energy at $r_{ms}$ at the end of the simulation.
Small errors in the gravitational and orbital energy can, in principle, then
lead to large errors in the other elements of the energy budget.  However,
to answer the questions of interest, an accurate accounting of magnetic,
thermal, and random kinetic energies is important.

In real disks, the energy of matter reaching the event horizon is
determined by a competition between several rates: of energy transfer
via work done on other matter and magnetic fields; of turbulent
dissipation; of magnetic reconnection; of photon creation; and
of photon diffusion.  The rate of energy transfer via work is
determined by the structure of the large-scale magnetic field and
fluid motions.  Turbulent dissipation is due to a two-part process:
nonlinear wave interactions transfer energy from long wavelengths to
short, where they can in turn be dissipated into heat by resistivity
and viscosity.  Magnetic reconnection is controlled by the rate at
which fluid motions force together fluid elements with oppositely
directed magnetic field.  When the gas is heated, a variety of
processes (bremsstrahlung, Compton scattering, etc.) transfer its heat
into photon energy.  Finally, energy leaves the disk as photons diffuse
away.

  Generally speaking, in these simulations we account reasonably
accurately for large-scale energy transfer, overall disk evolution, and
the long-wavelength portion of the turbulent cascade.  However,
numerical effects substitute for small-scale dissipation and enhance
reconnection.  Because the nonlinear wave interaction rate on long
lengthscales is very rapid compared to the numerical dissipation rate,
the effective turbulent dissipation rate in the simulation is controlled
by numerical effects.  Radiation we ignore entirely, with regard to
both photon creation and photon diffusion.  The absence of a real
treatment of dissipation and photon losses hinders our ability to
estimate the energy released by accretion.  For example, in HK01 we
argued that the observed change in energy between $R=3$ and $R_{min}$
was likely to be an underestimate because, in the absence of radiative
losses, some of the work done by plunging matter on gas farther out
would eventually be brought back in.

    However, at some level numerical effects mimic radiation by causing
energy to be lost when converging fluid elements carry with them
oppositely directed velocity or magnetic field.  We can estimate the
relative importance of numerical energy losses by a simple argument.
If energy were exactly conserved locally, the only change in the
integrated energy over the problem area would be due to material that
flows off the grid.  Here that effectively means material accreted
because, for example, over the course of the poloidal simulation
19\% of the initial mass passes
through the inner boundary while only 1.5\% leaves through the outer
boundary.  The binding energy of the accreted matter is, on average,
about 0.118 in our units, so that mass-loss alone should increase
the energy by 0.0224 if the initial mass is defined as unity.  Starting
from a total energy of -0.042, that means the final energy should be
-0.0196.  However, we actually find a final energy of -0.0272.  The
difference (a loss of 0.0076) can be attributed to numerical losses.
Because the numerical dissipation time is shorter than the inflow
time over much of the disk, this total energy loss is very close to
the total radiation loss predicted by the conventional model, but
there is no reason to believe that the specific location and rate
of numerical losses in the simulation accurately predict what would
happen in a real disk.

\subsection{Stress at the marginally stable orbit}

In these simulations we observe stress at the marginally stable orbit, yet
it has been a long-standing belief that the stress must be close to zero at
that point.  As discussed in \S1, at least two such arguments
have been put forward in support of this notion.
One is that the flow in the
plunging region would necessarily be low density and would lack the
inertia necessary to exert a torque on the disk (Novikov \& Thorne
1973; Page \& Thorne 1974).  However, Page \& Thorne point out that
this argument doesn't apply if the torques are carried by magnetic
fields.

The second argument relies upon the $\alpha$ stress model, i.e., that
the stress is $\alpha\rho c_s^2$.  This point is worth revisiting in
slightly more detail.  Equation (\ref{steadystress}) relates the mass
accretion rate to the stress in a steady state accretion flow.
Using the $\alpha$ model for the stress, this equation can be rewritten
\begin{equation}\label{alphastress}
\alpha_{SS} (c_s/v_R) (c_s/R\Omega) = \Delta l/l(R)
\end{equation}
where $\Delta l$ is the
difference between the local angular momentum $l(R)$ and the angular
momentum carried into the hole, $l(r_g)$.  Once the inflow passes
$r_{ms}$, the inflow speed increases and the pressure drops
precipitously, and with it the stress according to the $\alpha$
prescription.  The $\alpha$ model is implicitly {\it hydrodynamic},
so the point at which plunging matter loses contact with the disk is
the sonic point.  There, by definition, $v_R \sim c_s$, and
$c_s/R\Omega = H/R < 1$.  Since
$\alpha_{SS}$ is generally thought to be considerably less than unity,
it would seem to follow that $l$ changes very little in the plunging
region.  This is the basis for the ``proof'' that the stress goes to zero
at $r_{ms}$.

In the simulations, however, we find a scaled stress, i.e. $\Delta l/l$,
at $r_{ms}$ that is $\approx 0.05$ -- 0.1.  Why does the proof fail?
Because the reasoning
used relies on a strict application of the $\alpha$ relation, that is,
that the pressure {\it determines} the stress.  But the magnetic
stress in the plunging region doesn't depend upon the gas pressure;
$\alpha$ climbs rapidly inside $r_{ms}$ as the gas pressure drops and
the field evolves mainly by flux freezing.  In fact, we find in the
initially poloidal simulation that $\alpha_{SS}$ increases by a factor
of 100 as matter falls through this zone.

Since the torque is magnetic, can we save relation
(\ref{alphastress}) by replacing gas
pressure with magnetic pressure, and setting the torque equal to
$\alpha_{mag} \rho v_A^2$?  Even with this readjustment, $\alpha$
still doesn't determine the stress in a manner that can be used in
a proof; $\alpha_{mag}$ is also not a constant in the plunging
region.  It is roughly constant with value 0.2--0.3 in the disk
proper where magnetic turbulence is active, but rises toward 1 in the
plunging region.  The stress in the plunging region becomes more
efficient relative to the magnetic pressure as the radial field is
stretched out in the inflow, and its strength relative to the toroidal
field grows.  Moreover, $v_A/(R\Omega)$ increases slowly inward in
the flux-freezing regime, whereas in the conventional $\alpha$ model
$c_s/(R\Omega)$ falls rapidly inward.

These problems illustrate the difficulties that can arise when the $\alpha$
formalism is taken too literally.  The $\alpha$ parametrization cannot
be used as the basis of a proof.  At best, it might be possible to
use the reformulated version of this argument using $\alpha_{mag}$
to estimate a time-averaged $\Delta l/l$.   In that case, the observed
change in $l$ of 5--10\% inside the plunging region is
consistent with the value of $v_A/(R\Omega)$ that we find ($\simeq
0.15$ for the poloidal field case, $\simeq 0.05$ for the toroidal
field case).

\section{Conclusion}

In this paper we report on two new three-dimensional MHD global
accretion disk simulations, one beginning with a poloidal field and the
other with a purely toroidal field.  Since the topology of the magnetic
field in a real disk is largely unknown (except for the observation
that it is almost certainly dominated by the toroidal component) these
two initial configurations are intended to bracket the possible range.
We find that initial field topology makes a significant quantitative
difference in the resulting evolution.  With a purely toroidal field,
initial field amplification is slower, and the saturation energies,
magnetic stresses, and accretion rate are smaller, The magnetic stress
begins to evolve by flux freezing at a smaller radius in the toroidal
field run.  However, the qualitative features of angular momentum
transport and stress at the marginally stable orbit are unchanged.  In
any event, we expect that because the MRI is so effective at amplifying
poloidal field, real disks are likely to resemble the
initially-poloidal simulation more than the purely toroidal one.

We reexamined the issue of stress at the marginally stable orbit, and
confirmed our previous finding (HK01) that the stress remains
significant there.  The character of the stress changes as the flow
moves from turbulence to inflow.  Within the body of the disk, where
the dissipation time is short compared to the inflow time, the field
evolves by MRI-driven MHD turbulence limited by dissipation (numerical
in the simulation, resistive and viscous in real disks).  The relative
correlation of $B_R$ and $B_{\phi}$ ($\alpha_{mag}$) in that regime is
$\simeq 0.2$--0.3.  As the flow approaches $r_{ms}$ from the outside,
the inflow time falls until it is shorter than the dissipation time.
Inside that radius, the field evolves mainly via flux freezing and
$\alpha_{mag}$ increases toward unity.  The location of this transition
is one of the most important parameters governing the character of the
inner accretion flow.  As a result of the continuing stress in the
plunging region, inside $r_{ms}$ accreting matter suffers a modest
($\sim$10\%) decline in its average  specific angular momentum and a
comparable increase in its binding energy.  Rapid variability in
accretion rate remains the rule with the most power at frequencies
lower than the orbital frequency at $r_{ms}$.

These simulations used the largest number of grid zones and arranged
them so as to achieve the finest resolution in the inner accretion flow
of any simulation to date.  We find that the increased resolution used
in these simulations produced increased field energy and stress, and
also lead to larger amplitude variations in the accretion rate.  We
believe that at this level of resolution we are approaching numerical
convergence with respect to the magnetic field intensity and the
accretion rate for the initially-poloidal case, but finer resolution
may be necessary to demonstrate convergence for these quantities when
the initial field topology is toroidal, and for other quantities for
both topologies.  Unfortunately, it is difficult to obtain
significantly greater resolution or to use this high resolution over a
wider range of radius.  Certain questions will therefore remain
difficult to answer for some period of time (most importantly, the
location of the dissipation/flux-freezing transition radius).  Perhaps
this is not surprising, considering that we are attempting to resolve a
turbulent cascade within a global disk.  On the other hand, the overall
evolution and qualitative features have remained consistent from the
lowest to highest resolutions.  This is fortunate since the dynamics of
self-consistent MHD accretion flows remain largely unexamined, and
further exploration with even the modest resolution permitted by
current practical considerations is likely to be fruitful.

\begin{acknowledgments}

This work was supported by NSF grant AST-0070979, and NASA grants
NAG5-9266 and NAG5-7500 to JFH, and NASA grant NAG5-9187 to JHK.
Simulations were carried out on Bluehorizon, the IBM SP cluster of the
San Diego Supercomputer Center of the National Partnership for Advanced
Computational Infrastructure, funded by the NSF, and on Centurion, a
linux-based alpha cluster operated by the Legion project of the
University of Virginia Computer Science Department, Andrew Grimshaw
PI.

\end{acknowledgments}

\end{document}